\documentclass[twocolumn]{aastex63}

\usepackage{amsmath} 

\shorttitle{What makes quads quadruple?}
\shortauthors{Luhtaru, Schechter \& de Soto}

\submitjournal{ApJ}
\accepted{April 28, 2021}

\begin{document}
	
	\title{What Makes Quadruply Lensed Quasars Quadruple?}
	
	\correspondingauthor{Paul L. Schechter}
	\email{schech@achernar.mit.edu}
	
	\author[0000-0003-2187-7090]{Richard Luhtaru}
	\affiliation{MIT Department of Physics \\
		Cambridge, MA 02139, USA}
	
	\author[0000-0002-5665-4172]{Paul L. Schechter}
	\affiliation{MIT Department of Physics \\
		Cambridge, MA 02139, USA}
	\affiliation{MIT Kavli Institute for Astrophysics and Space Research \\
		Cambridge, MA 02139, USA}
	
	\author[0000-0002-9886-2834]{Kaylee M. de Soto}
	\affiliation{MIT Department of Physics \\
		Cambridge, MA 02139, USA}
	
	\begin{abstract}
		
		Among known strongly lensed quasar systems, $\sim$25\% have gravitational potentials sufficiently flat \added{(and sources sufficiently well aligned)} to produce four images rather than two. The projected flattening of the lensing galaxy and tides from neighboring galaxies both contribute to the potential's quadrupole. Witt's hyperbola and Wynne's ellipse permit determination of the overall quadrupole from the positions of the quasar images. The position of the lensing galaxy resolves the distinct contributions of intrinsic ellipticity and tidal shear to that quadrupole. Among 31 quadruply lensed quasars systems with statistically significant decompositions, 15 are either reliably ($2\sigma$) or provisionally ($1\sigma$) shear-dominated and 11 are either reliably or provisionally ellipticity-dominated. For the remaining 8, the two effects make roughly equal contributions to the combined cross section (newly derived here) for quadruple lensing. This observational result is strongly at variance with the ellipticity-dominated forecast of \citet{oguri}.
		
	\end{abstract}
	
	
	
	\section{Introduction} \label{sec:intro}
	
	Lensed quasars can be used to study the quasars themselves, the lensing galaxies, and the intervening and associated intergalactic medium \citep{zahedy}. The best-known use of strongly lensed quasars is measuring the Hubble parameter $H_0$ \citep{refsdal,treumarshall}. While the many applications of gravitational lensing have long been known, their systematic exploitation has been relatively recent, enabled by the advent of multiple wide field surveys, most notably Gaia \citep{delchambre}.
	
	A crucial prerequisite for drawing conclusions from lensed quasars is a model for the mass of the lens. There are many possible models and multiple degeneracies among them \citep{schneidersluse} that preclude determination of a unique projected two-dimensional gravitational potential using only four quasar images.
	
	In this paper, we explore one such degeneracy, \added{first identified by \citet{kassiolakovner},} between external shear and ellipticity, in a restricted variant of the singular isothermal elliptical potential (SIEP+XS). It is a close cousin of the singular isothermal elliptical mass (SIE+XS) most often used in literature \citep[e.g.][]{kks}. We assume that ellipticity and shear are aligned in the same direction (possibly with different signs). It is a \textit{virtue} of this model that shear and ellipticity are completely degenerate if it is constrained by only the four image positions of a quadruply lensed quasar (henceforth a ``quad''). We show that adding an additional constraint, the measured position of the lensing galaxy, breaks the degeneracy. We then use the measured positions for a sample of single-lens quads to do just this.
	
	Shear and ellipticity are the two principal sources of the asymmetry needed for a quadruply lensed quasar. While a singular isothermal sphere (SIS) with no external shear produces only two images, increasing ellipticity or shear increases the probability of having four images \citep{huterer}. Although it is also possible to have ``naked cusps'' with only three images in cases of very high shear ($\gamma > 1/3$), these configurations are known to be very rare among systems discovered so far \citep{finch}. By breaking the degeneracy of shear and ellipticity, we can ascertain, for a given system, which of them plays the larger role in producing four images.
	
	Until now the samples available for such an analysis have been small. \replaced{\citet{wong} analyze 6 quads and \citet{shajib} analyze 13. Neither paper}{\citet{kks}, \citet{wong}, and \citet{shajib} analyze, respectively, 4, 6, and 13 quads. None of these papers} explicitly addresses the question, but the precepts described below can be used to determine the relative contributions of ellipticity and shear. The contributions are roughly equal for the \citet{wong} sample but the ellipticities contribute somewhat more in the \citet{shajib} sample. Most of the lenses from both samples are included in the uniform analysis below.
	
	In Section \ref{sec:background}, we give an analytic description of the SIEP+XS model, report the main results of \citet{witt1996} and \citet{wynne}, and explain the degeneracy between shear and ellipticity and how they relate to the probability of producing four images. In Section \ref{sec:method}, we explain how we use the model and observed positions of lensing galaxies to estimate the dominant factor in each system. In Section \ref{sec:results}, we describe the results and estimate errors of this method for known single-lens systems. In Section \ref{sec:discussion}, we test our method with a simulated mock \replaced{catalogue}{catalog} created by \citet{oguri} and compare the decompositions obtained on observed systems with the simulated systems.
	
	
	\section{Background} \label{sec:background}
	
	\subsection{Restricted SIEP+XS model}
	
	The model we use for the lensing potential is \explain{removed unnecessary line break}
	\begin{multline}\label{eq:psi}
		\psi(x,y) = b\sqrt{q_\text{pot}\left(x-x_{g}\right)^2 + \frac{\left(y-y_g\right)^2}{q_\text{pot}}} \\- \frac{\gamma}{2}\left(\left(x-x_s\right)^2 - \left(y-y_s\right)^2\right),
	\end{multline}
	where $\mathbf{r} = (x,y)$ is the position in the plane of the sky in a frame aligned with the axes of the potential, $(x_g, y_g)$ is the position of the lensing galaxy, $(x_s, y_s)$ is the true position of the source (quasar), $b$ characterizes the lens strength, $q_\text{pot}$ is axis ratio of the potential, and $\gamma$ is shear\footnote{Note that shear is centered on the source and not on the galaxy in our model. One model can be obtained from the other by adding a wedge potential, which shifts the images but does not change the relative positions \citep{gorenstein}.}. $\psi$ is the conventional definition of the lens potential:
	\begin{equation}
		\psi(\mathbf{r}) = \frac{D_{ls}}{D_l D_s}\frac{2}{c^2}\int_0^\infty \Phi\left(D_l \mathbf{r}, z\right)\,\mathrm{d}z,
	\end{equation}
	where $\Phi$ is the Newtonian gravitational potential of the lens, $z$ is the line-of-sight coordinate, and $D_l$, $D_s$, $D_{ls}$ are respectively the angular-diameter distances to the lens, to the source, and between the lens and the source, and
	\begin{equation}\label{eq:lensequation}
		\mathbf{r} - \mathbf{r}_s = \boldsymbol{\nabla}\psi(\mathbf{r})
	\end{equation}
	is the deflection of the image \citep{bourassa1975}.
	
	This model has two interesting geometric properties:
	\begin{enumerate}
		\item All four image positions, the source, and the center of the lensing galaxy lie on a rectangular hyperbola, described by the equation
		\begin{equation}\label{eq:hyperbola}
			\frac{y-y_s}{x-x_s} = \left(\frac{1+\gamma}{1-\gamma}\right)\frac{1}{q_\text{pot}^2}\frac{y-y_g}{x-x_g}.
		\end{equation}
		
		\item All four images a) lie on an ellipse that is b) aligned with the asymptotes of hyperbola and c) centered on the source d) which itself lies on the hyperbola, described by the equation
		\begin{equation}\label{eq:ellipse}
			\left(x-x_s\right)^2 + \left(\frac{1-\gamma}{1+\gamma}\right)^2 q_\text{pot}^2 (y-y_s)^2 = \frac{b^2q_\text{pot}}{\left(1+\gamma\right)^2}.
		\end{equation}
	\end{enumerate}
	
\begin{figure*}[htb!]
		\gridline{\fig{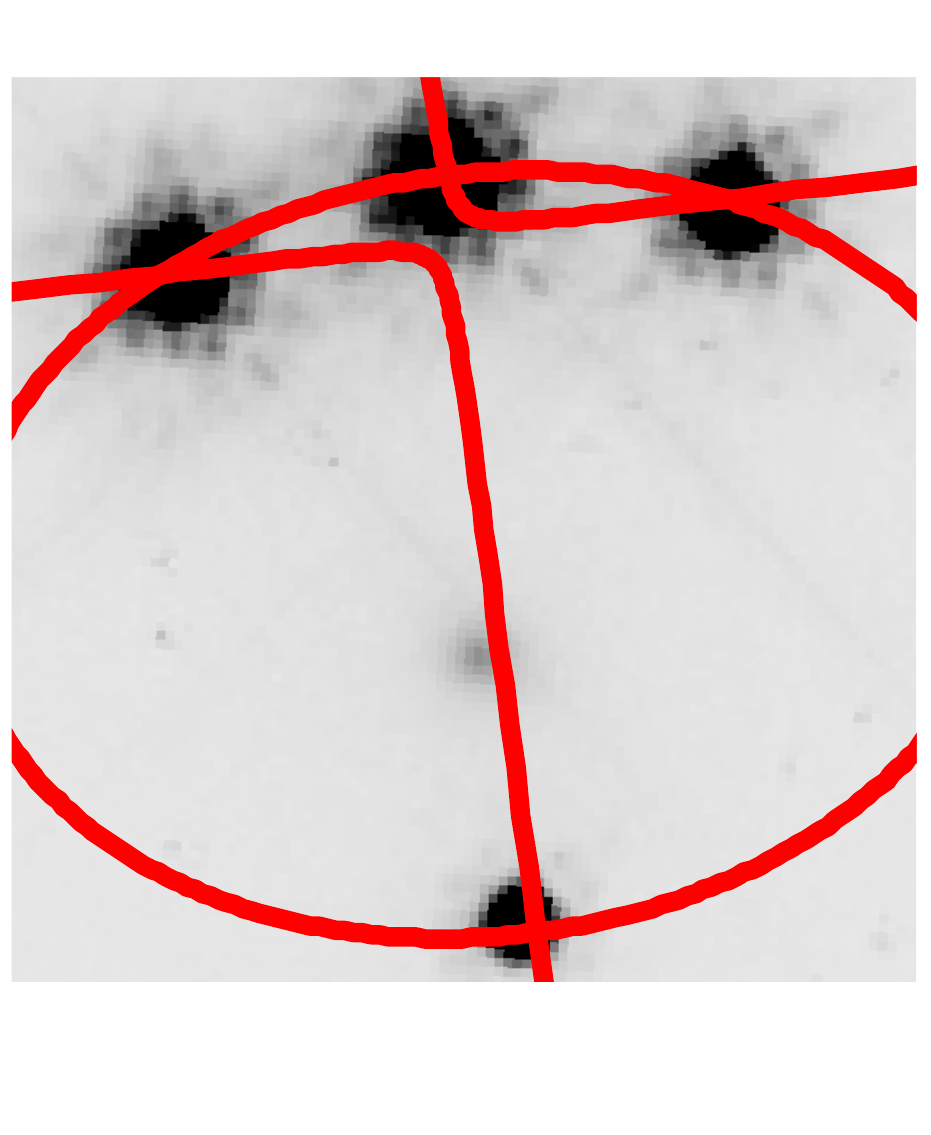}{0.348\textwidth}{(a)}
			\fig{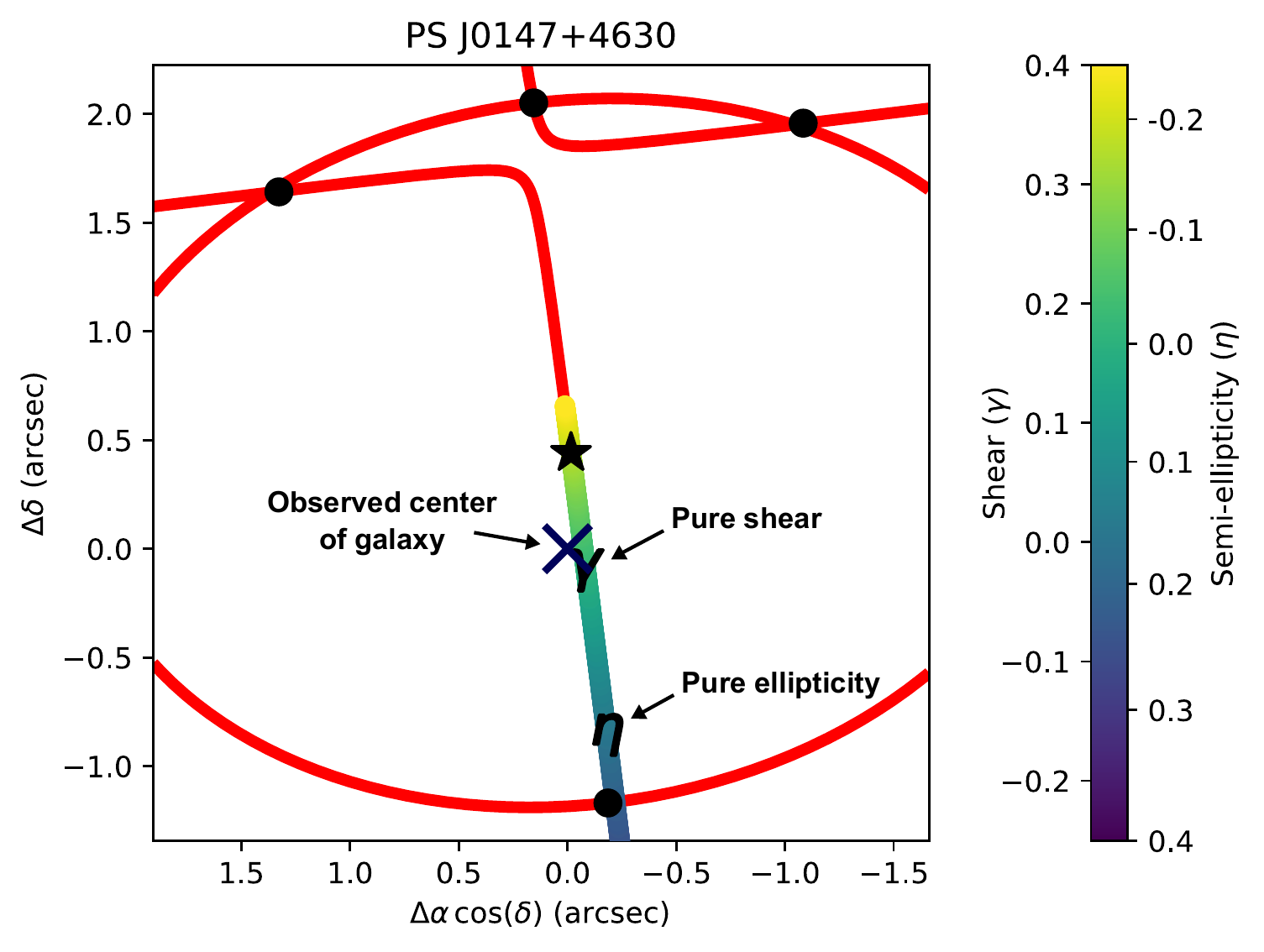}{0.56\textwidth}{(b)}}
		\caption{(a) Image positions of system PS J0147+4630 and the best-fit hyperbola and ellipse passing through them. (b) Predicted positions of the galaxy center of system PS J0147+4630 for different values of shear \added{and ellipticity}. $\gamma$ indicates \added{the} position for pure shear ($\eta = 0$), $\eta$ indicates \added{the} position for pure ellipticity ($\gamma = 0)$, the cross marker indicates the observed position, the star marker indicates the predicted position of the source. \added{The offset from the hyperbola is discussed further in Section \ref{sec:method}.}\label{fig:galpos}}
	\end{figure*}

	The first property is shown by \citet{witt1996} \added{to hold} for both the SIEP and SIS+XS \replaced{models and the}{models.\footnote{\citet{zhaopronk} describe a broader class of quasi-elliptical models that can be fit by a ``semi-hyperbolic'' curve.} The} second property is shown by \citet{wynne} \added{to hold} for a SIEP model; \replaced{they are also}{both properties also hold} true for our model (as we derive in Appendix \ref{app:geom}). An example of the two properties is shown in Figure \ref{fig:galpos}a.
	
	\explain{Changed color bar of Figure 1b to be two-sided.}
	
	\subsection{Degeneracy of shear and ellipticity}
	
	As can be seen from the equation of the second property, the axis ratio of the ellipse passing through the four images (which we will call the ``deflection'' ellipse) is
	\begin{equation} \label{eq:qell}
		q_\text{ell} = \left(\frac{1-\gamma}{1+\gamma}\right) q_\text{pot}.
	\end{equation}
	\deleted{As shown in Appendix \ref{app:imageequivalence}, the same four-image configuration may be produced only with pure shear or an equivalent amount of pure ellipticity (or any combination of them, given that $q_\text{ell}$ is the same).} To handle shear and ellipticity symmetrically, we define \textit{semi-ellipticity} $\eta$ so that $q_\text{pot} = (1-\eta)/(1+\eta)$. It follows that
	\begin{equation}
		\eta \equiv \frac{1-q_\text{pot}}{1+q_\text{pot}} = \frac{e_\text{pot}}{2-e_\text{pot}},
	\end{equation}
	where $e_\text{pot} = 1 - q_\text{pot}$ is the conventional definition of ellipticity (flattening). Note that for small values of ellipticity, $\eta \approx e/2$. We have also implicitly assumed that $\gamma$ and $\eta$ are parallel; otherwise, they denote components parallel to the effective quadrupole.
	
	Equation (\ref{eq:qell}) now rewrites as
	\begin{equation}
		q_\text{ell} = \left(\frac{1-\gamma}{1+\gamma}\right)\left( \frac{1-\eta}{1+\eta}\right),
	\end{equation}
	which makes it clear that a given \replaced{quadruple}{deflection ellipse} can be produced by an equal value of pure shear $\gamma$ or an equal value of pure semi-ellipticity $\eta$ as they produce the same flattening of the \deleted{deflection }ellipse.
	
	We also define \textit{effective quadrupole} $\Gamma_\text{eff}$ so that
	\begin{equation}
		q_\text{ell} = \frac{1-\Gamma_\text{eff}}{1+\Gamma_\text{eff}},
	\end{equation}
	which is the value of shear (or semi-ellipticity) needed to produce a given \replaced{quadruple}{deflection ellipse} if it was a system with pure shear (or pure ellipticity). Introducing semi-ellipticity provides an easy way to determine the dominant cause of asymmetry (flattening of the deflection ellipse) in a given system -- it is the one with a higher absolute value. Note that as we define $q_\text{ell} < 1$, the parameter with a higher absolute value must be positive.
	
	\added{If one similarly substitutes the semi-ellipticity for the axis ratio in equation (\ref{eq:hyperbola}) for our hyperbola, one finds that shear and semi-ellipticity do \textit{not} enter in the same functional form. We show in Appendix \ref{app:imageequivalence}, that the \textit{identical hyperbola} is recovered if we simultaneously change the galaxy position.} Paralleling the development in \citet{wynne}, we show that in the aligned frame with the origin at the center of the hyperbola, \added{the position of the galaxy is given by}
	\begin{equation} \label{eq:xgal}
		x_g = \left(\frac{1+\gamma}{1-\gamma}\right)\frac{1}{q_\text{pot}^2} x_s
	\end{equation}
	and
	\begin{equation} \label{eq:ygal}
		y_g = \left(\frac{1-\gamma}{1+\gamma}\right) q_\text{pot}^2 y_s.
	\end{equation}
	The same four-image configuration may \added{therefore} be produced\deleted{ only} with pure shear\replaced{ or}{,} an equivalent amount of pure ellipticity\replaced{ (}{, }or any combination of them, given that $q_\text{ell}$ is the same\deleted{)}.
	
	\added{A number of investigators have previously reported variants and generalizations of this result. \citet{kassiolakovner} have found that for any assumed circular potential with external shear, there is an elliptical potential of the same functional form with no shear that leaves ``unchanged the qualitative properties of the lensing'', as do all members of a one-dimensional intermediate family. They define a $\gamma^2_\text{inv}$ that reduces to our $\Gamma_\text{eff}$ for the case of aligned shear and no external convergence.}
	
	\added{A variant of the \citet{kassiolakovner} result was rediscovered by \citet{wittmao}, who noted that the lens equation for an elliptical power law potential with an aligned external shear can be transformed into the lens equation for an equivalent pure elliptical potential. \citet{jinan} finds that a singular isothermal elliptical mass distribution (SIE) with an external shear can be modeled with an equivalent pure SIE ``provided that the deviation from circular symmetry is small'' and goes on to say that this can be generalized to yielding ``a certain degeneracy'' involving the shear, ellipticity and
		their relative orientations.}
	
	\added{In Appendix \ref{app:imageequivalence} we explicitly exhibit the one-dimensional family of models, including the SIEP ($\gamma = 0$), the SIS+XS ($q_\text{pot} = 1$), and the range of models in between (and beyond) that produce identical image configurations with a common $q_\text{ell}$.}
	
	\subsection{Probability of having four images}
	
	Although $\gamma$ and $\eta$ are symmetric in terms of the flattening, they do not have symmetric contributions to the probability of producing four images. Assuming a random distribution of visible quasars in the sky \added{and neglecting magnification bias (which is discussed in Section \ref{sec:magbias})}, the probability of a given lens having a quadruply lensed quasar is \deleted{directly }proportional to the area of the \replaced{astroid}{astroidal} caustic -- the region in the sky where a source would be quadruply lensed \citep{finch}. As derived in Appendix \ref{app:caustic}, the area of the \replaced{astroid}{astroidal} caustic of a lens potential described by equation (\ref{eq:psi}) is
	\begin{equation}
		A = \frac{3\pi b^2 \left(\left(1-q_\text{pot}^2\right)+\left(1+q_\text{pot}^2\right)\gamma\right)^2}{8(1-\gamma^2)q_\text{pot}^2}
	\end{equation}
	
	\added{This result is consistent with the result of \citet{jinan} for an SIE with external shear}. As a lowest-order approximation with $|\gamma| \ll 1$ and $|\eta| \ll 1$ we have $q_\text{pot} \approx 1-2\eta$ and
	\begin{equation}
		A \approx \frac{3\pi b^2}{2}\left(\gamma + 2\eta\right)^2
	\end{equation}
	Therefore, in the limit of small $\gamma$ and $\eta$, shear and ellipticity have equal contributions to the probability of producing a quadruplet if $|\gamma| = |2\eta|$.
	
	One way to measure the contributions of $\gamma$ and $\eta$ in the general case is to compare the values of $A(\gamma, \eta=0)$ and $A(\gamma=0, \eta)$. Knowing that $q_\text{pot} = (1-\eta)/(1+\eta)$, we obtain
	\begin{align}
		A_\gamma \equiv A(\gamma, \eta=0) &= \frac{3\pi b^2}{2}\frac{\gamma^2}{1-\gamma^2}\\
		A_\eta \equiv A(\gamma=0, \eta) &= \frac{3\pi b^2}{2}\frac{4\eta^2}{\left(1-\eta^2\right)^2}
	\end{align}
	
	Henceforth, we will call a system with known $\gamma$ and $\eta$ values \textit{shear-dominated} if $A_\gamma > A_\eta$ and \textit{ellipticity-dominated} if $A_\gamma < A_\eta$.
	
	\section{Methodology} \label{sec:method}
	
	\explain{Moved equations (\ref{eq:xgal}) and (\ref{eq:ygal}) to Section 2.2}
	
	As explained in the previous section, it is not possible to differentiate external shear and ellipticity of the potential in our model using only the four image positions\replaced{. However, the galaxy position needed to produce a given configuration is not the same for different proportions of shear and ellipticity.}{, but they can be differentiated by additionally knowing the galaxy position.} Because $x_s$ and $y_s$ are fixed and uniquely determined by the rectangular hyperbola and the deflection ellipse, \replaced{with}{and} the coefficient \replaced{this time}{is} proportional to $q_\text{pot}^2$ \added{in equations (\ref{eq:xgal}) and (\ref{eq:ygal})} as opposed to $q_\text{pot}$ in equation (\ref{eq:qell}), $x_g$ and $y_g$ are not fixed, even if $q_\text{ell}$ stays fixed.
	
	Therefore, if we also know the position of the lensing galaxy, we can estimate the proportion of shear and ellipticity by comparing the observed galaxy position with our model and choosing the proportion that gets the modeled galaxy position closest to it. An example is shown in Figure \ref{fig:galpos}b, where the system appears to be shear-dominated according to our method.
	
	We use this method for 39 known quadruply lensed quasar systems with a single lensing galaxy that have accurate data for the positions of the four images and the position of the galaxy, listed in Table \ref{tab:systems} in Appendix \ref{app:positions}. We have excluded systems with multiple lensing galaxies because our model assumes only one lens.
	
	To gauge the uncertainty in our estimated proportions of shear and ellipticity, we use both the distance between the true (observed) galaxy position and the hyperbola, as well as the published galaxy position uncertainty, which is especially important for systems with small $\Gamma_\text{eff}$, where the estimated proportion is very sensitive to change in position and having the true position randomly aligned with the hyperbola could lead to a significant underestimation of the uncertainty. We \added{conservatively} take
	\begin{equation}
		\Delta \theta = \Delta \theta_d + \Delta \theta_p,
	\end{equation}
	where $\Delta \theta_d$ is the angular distance between the true and the best derived galaxy position and $\Delta \theta_p$ is the published uncertainty of the true galaxy position (maximum of $\alpha \cos \delta$ and $\delta$ uncertainty). We then consider a possible predicted galaxy position to be in $1\sigma$ range if its distance from the best derived position is less than $\Delta \theta$, and we estimate $\Delta \gamma$ to be half of the range of the $\gamma$ values in the $1\sigma$ range. Similarly, we consider points on the hyperbola closer than $2\Delta \theta$ to be in $2\sigma$ range.
	
	In summary, our method is the following:
	\begin{enumerate}
		\item Find the best-fit hyperbola and deflection ellipse passing through the four images (explained in more detail in Appendix \ref{app:fitting}).
		\item Use the best fit to calculate predicted positions of the galaxy for different shear and ellipticity decompositions in the frame aligned with the asymptotes of the hyperbola using equations (\ref{eq:xgal}) and (\ref{eq:ygal}).
		\item Convert the predicted positions back to the observed frame and find the one closest to the observed position of the galaxy to determine the best decomposition.
		\item Use the published galaxy position uncertainty and the distance of the observed and the closest derived position to estimate the uncertainty of our method.
	\end{enumerate}
	
	\explain{This sentence might be misleading without context, because we minimized distances to hyperbola/ellipse, not to intersection points. If we minimized distances to intersection points, then higher weights are possibly not needed.}
	\deleted{When fitting the hyperbola and ellipse through the images, we used higher weights for images closer to each other because they gave significantly better fits for systems with two very close images (see appendix \ref{app:fitting}).}
	
	
	\section{Results} \label{sec:results}
	
\begin{figure*}[htb!]
		\gridline{\fig{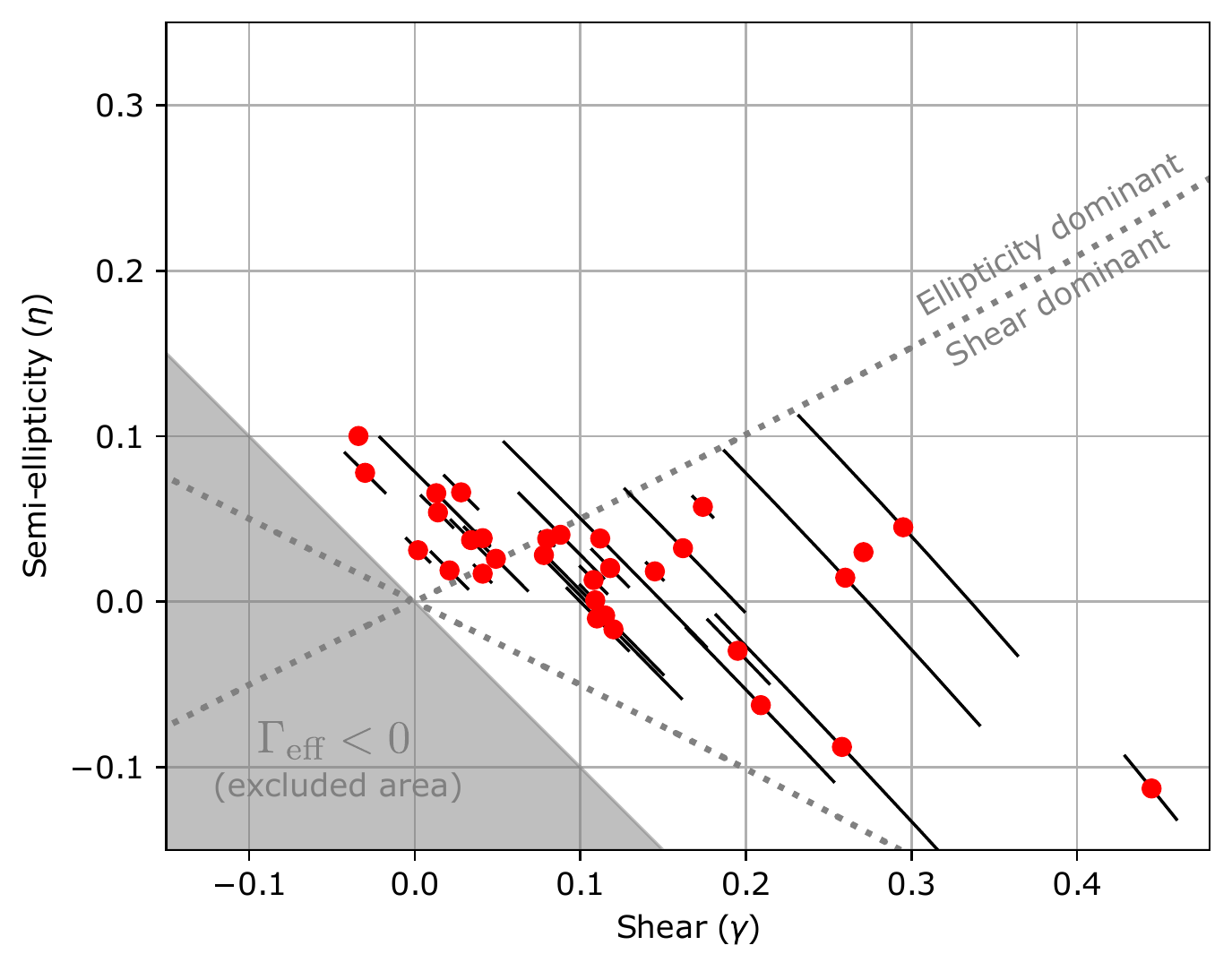}{0.47\textwidth}{(a)}
			\fig{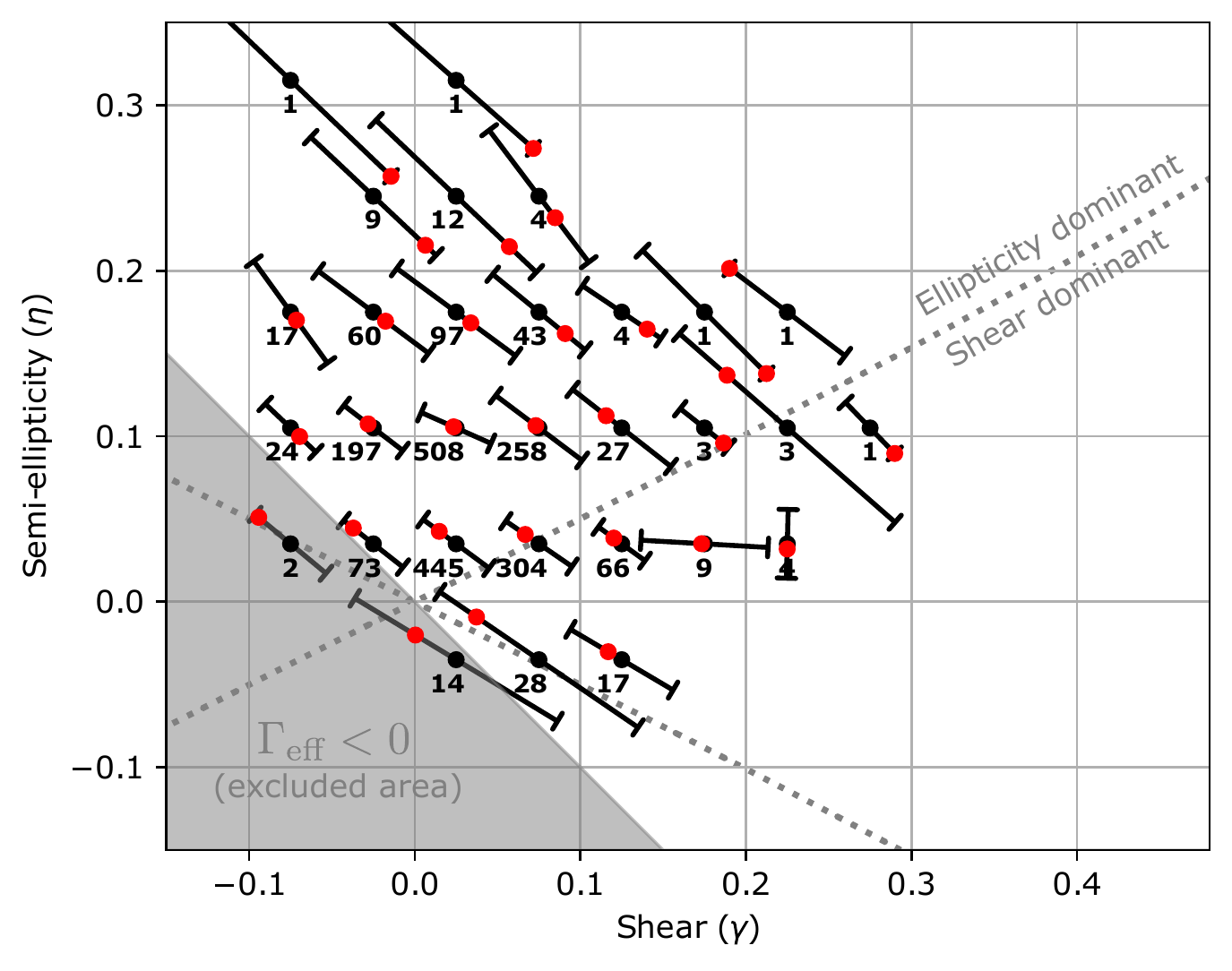}{0.47\textwidth}{(b)}}
		\caption{(a) Estimated shear and ellipticity values of 31 observed single-lens systems with an acceptably low modeling error ($\Delta \gamma / \Gamma_\text{eff} < 0.5$). Black bars indicate estimated $1\sigma$ uncertainty range. (b) Comparison of true and estimated shear and ellipticity components (parallel to the effective quadrupole) of 2233 simulated quadruply lensed quasar systems in the mock \replaced{catalogue}{catalog} of \citet{oguri}. Results have been averaged by dividing the graph into boxes. Black dots and the corresponding numbers show the center and the number of systems in each box. Red dots show the mean estimated decomposition. Their offsets from the black dots show the systematic error. Black bars show the random error (RMS difference). The dotted gray line indicates points where ellipticity and shear contribute equally to the probability of producing four images ($A_\gamma = A_\eta$). \label{fig:gammavseta}}
	\end{figure*}

	The estimated \replaced{proportions}{shear and ellipticity decompositions} for 31 out of 39 systems with a low enough modeling error ($\Delta \gamma / \Gamma_\text{eff} < 0.5$) are shown in Figure \ref{fig:gammavseta}a. The signs of $\eta$ and $\gamma$ are chosen so that $q_\text{ell} < 1$ (i.e. $\Gamma_\text{eff} > 0$).
	
	We conclude that out of the 39 systems:
	\begin{itemize}
		\item \textbf{10 are reliably shear-dominated}\\
		($A_\gamma > A_\eta$ for every position in $2\sigma$ range)
		\item \textbf{5 are provisionally shear-dominated}\\
		($A_\gamma > A_\eta$ for every position in $1\sigma$ range)
		\item \textbf{13 are uncertain}\\
		($A_\gamma = A_\eta$ for some position in $1\sigma$ range)
		\item \textbf{5 are provisionally ellipticity-dominated}\\
		($A_\eta > A_\gamma$ for every position in $1\sigma$ range)
		\item \textbf{6 are reliably ellipticity-dominated}\\
		($A_\eta > A_\gamma$ for every position in $2\sigma$ range)
	\end{itemize}
	
	Estimated \replaced{proportions}{decompositions} for all 39 systems are given in Table \ref{tab:results}.
	
	\startlongtable
	\begin{deluxetable}{lDDDD}
		\tablecaption{Estimated shear and ellipticity values of all 39 analyzed single-lens systems \label{tab:results}}
		\tablehead{
			\colhead{System name} & \twocolhead{$\gamma$} & \twocolhead{$\eta$} & \twocolhead{$\Delta \gamma$} & \twocolhead{$\Gamma_\text{eff}$}
		}
		\decimals
		\startdata
		J0029-3814 & 0.445 & -0.113 & 0.016 & 0.350 \\
		J0030-1525 & -0.034 & 0.100 & 0.002 & 0.066 \\ 
		PS J0147+4630 & 0.195 & -0.030 & 0.018 & 0.166 \\ 
		SDSS J0248+1913 & 0.120 & -0.017 & 0.043 & 0.104 \\ 
		ATLAS J0259-1635 & 0.014 & 0.054 & 0.010 & 0.068 \\ 
		DES J0405-3308 & -0.012 & 0.032 & 0.043 & 0.020 \\ 
		DES J0420-4037 & 0.021 & 0.019 & 0.011 & 0.040 \\ 
		HE0435-1223 & 0.049 & 0.026 & 0.019 & 0.075 \\ 
		J0530-3730 & 0.088 & 0.040 & 0.026 & 0.128 \\ 
		J0659+1629 & 0.013 & 0.066 & 0.033 & 0.079 \\ 
		B0712+472 & 0.078 & 0.028 & 0.003 & 0.106 \\ 
		HS0810+2554 & 0.011 & 0.008 & 0.051 & 0.019 \\ 
		RXJ0911+0551 & 0.271 & 0.030 & 0.003 & 0.299 \\ 
		SDSS0924+0219 & 0.041 & 0.017 & 0.005 & 0.058 \\ 
		J1042+1641 & -0.028 & 0.056 & 0.044 & 0.028 \\ 
		PG1115+080 & 0.109 & 0.001 & 0.009 & 0.110 \\ 
		RXJ1131-1231 & 0.118 & 0.020 & 0.011 & 0.138 \\ 
		J1131-4419 & 0.002 & 0.031 & 0.007 & 0.033 \\ 
		J1134-2103 & 0.295 & 0.045 & 0.066 & 0.336 \\ 
		SDSS1138+0314 & 0.110 & -0.010 & 0.019 & 0.100 \\ 
		SDSS J1251+2935 & 0.080 & 0.038 & 0.004 & 0.118 \\ 
		HST12531-2914 & 0.258 & -0.088 & 0.081 & 0.174 \\ 
		SDSS J1330+1810 & -0.030 & 0.078 & 0.012 & 0.048 \\ 
		HST14113+5211 & 0.260 & 0.015 & 0.077 & 0.273 \\ 
		H1413+117 & 0.137 & -0.027 & 0.155 & 0.110 \\ 
		HST14176+5226 & 0.162 & 0.000 & 0.144 & 0.162 \\ 
		B1422+231 & 0.174 & 0.057 & 0.006 & 0.229 \\ 
		SDSS J1433+6007 & 0.162 & 0.032 & 0.036 & 0.193 \\ 
		J1537-3010 & 0.209 & -0.062 & 0.045 & 0.148 \\ 
		PS J1606-2333 & 0.240 & -0.034 & 0.154 & 0.207 \\ 
		J1721+8842 & 0.108 & 0.013 & 0.008 & 0.121 \\ 
		J1817+27 & -0.095 & 0.120 & 0.091 & 0.025 \\ 
		WFI2026-4536 & 0.115 & -0.008 & 0.037 & 0.107 \\ 
		DES J2038-4008 & 0.028 & 0.066 & 0.010 & 0.094 \\ 
		B2045+265 & 0.145 & 0.018 & 0.005 & 0.163 \\ 
		J2100-4452 & 0.034 & 0.037 & 0.012 & 0.071 \\ 
		J2145+6345 & 0.112 & 0.038 & 0.061 & 0.150 \\ 
		J2205-3727 & 0.041 & 0.038 & 0.004 & 0.079 \\ 
		WISE J2344-3056 & -0.163 & 0.229 & 0.147 & 0.069 \\
		\enddata
	\end{deluxetable}
	
	
	\explain{Replaced old section title Discussion}
	\section{Comparison with simulated catalog} \label{sec:discussion}
	
	To test our method and to compare \replaced{the results with what is currently known about lensing galaxies}{our results with the current best estimates of lens properties}, we also use our algorithm to estimate the shear and ellipticity components of simulated quadruply lensed quasar systems in the mock \replaced{catalogue}{catalog} created by \citet{oguri}, which predicts the distribution of such systems in future optical imaging surveys. The mock \replaced{catalogue}{catalog} includes 2233 quadruplets with known image positions and shear and ellipticity values.
	
	Because the \replaced{catalogue}{catalog} uses an elliptical mass density model for the galaxies instead of elliptical lensing potential (which is a good approximation for small ellipticities), we first convert the axis ratio of mass density to axis ratio of lensing potential using the relation
	\begin{equation}
		q_\text{pot} = \frac{\tan^{-1}\left(\sqrt{1-q_\text{mass}^2}/q_\text{mass}\right)}{\tanh^{-1}\left(\sqrt{1-q_\text{mass}^2}\right)},
	\end{equation}
	which is discussed in more detail in Appendix \replaced{$\ref{app:potmassrelation}$}{\ref{app:potmassrelation}}.
	
	Additionally, we only compare the components parallel to the effective quadrupole with our estimates since our model assumes that shear and ellipticity are aligned. If the real shear and semi-ellipticity values are $\gamma_\text{tot}$ and $\eta_\text{tot}$ and the angle between them is $\phi$, we first find the direction of the effective quadrupole by adding two vectors with the same lengths in a double-angled space (where the angle between the two vectors is $2\phi$). We then find the parallel components $\gamma$ and $\eta$ by calculating the projections of the two individual vectors on that direction in the same space.
	
	After converting $\gamma_\text{tot}$ and $q_\text{mass}$ to $\gamma$ and $\eta$, we compare these values with the estimates from our algorithm based on the image and galaxy positions. The deviations of predicted and estimated true values are shown in Figure \ref{fig:gammavseta}b, where red dots indicate the systematic error and black bars indicate the random error.
	
	We see that the estimated values are generally consistent with the true values, especially for systems with small ellipticity. However, we notice that there is a significant discrepancy between the estimated shear and ellipticity values of observed systems and systems in the mock \replaced{catalogue}{catalog}. We estimate 62\% of the observed systems to be shear-dominated compared to only 10\% of the systems in the mock \replaced{catalogue}{catalog}, which suggests that shear is a more important factor in quadruply lensed systems than has been previously assumed. Even if our method is subject to systematic error, the systematic difference in estimates between observed and simulated systems implies that there are shortcomings in the ranges of ellipticities and shears used by \citet{oguri} to create their mock \replaced{catalogue}{catalog}. We note that \citet{collett} and \citet{goldstein} have used similarly narrow ranges of shears in forecasting the rates of lensed supernovae. The observed range of shears is more nearly consistent with those calculated by \citet{holderschechter} using N-body simulations.
	
	In our calculation of the effective quadrupole for the \citet{oguri} \replaced{catalogue}{catalog}, we combine shear and semi-ellipticity as if vectors in our double-angled space. Figure \ref{fig:gammavseta}b shows that the systematic errors in the relative contributions of shear and ellipticity to the effective quadrupole are small. The perpendicular components of shear and semi-ellipticity affect the image positions differently, but for the extreme case with $\gamma = \eta$ at $45^{\circ}$ to each other, those non-cancelling perpendicular components are only half the size of effective shear. For the \citet{shajib} sample, the average difference between the total shear and its component projected onto the effective shear \added{is} $0.015$. The corresponding average for the semi-ellipticity is $0.010$.
	
	While we attribute the difference between Figures \ref{fig:gammavseta}b and \ref{fig:gammavseta}a to errors in the underlying intrinsic shear and ellipticity distributions adopted by \citet{oguri}, one must ask whether differential selection effects might contribute to that difference.
	
	\replaced{The mock lens catalogue has a magnitude limit on the third brightest lensed image and a minimum separation between the two most distant images. \citet{finch} show that the mean magnification for purely sheared systems is roughly a factor of two larger than for purely flattened systems. We suspect that this also applies to the individual images for a given hyperbola and ellipse. The unmagnified limiting magnitude for a pure shear system is therefore $0.75$ magnitudes fainter than for a pure ellipticity system. But the area of a pure shear caustic (and hence the cross section for quadruple lensing) is one quarter that of a pure ellipticity caustic. Therefore, the surface density of quasars must increase as $10^{0.8m}$ -- much more rapidly than observed at the relevant apparent magnitudes -- for the two effects to cancel. The mock catalogue therefore favors flattened systems.
		
		The observed sample is drawn from many different surveys, each with different magnitude and separation criteria, but the flux from the third brightest image and the largest separation (or perhaps the larger distance of the two brighter images from the third) are likely to have played a similarly dominant role in their selection. The observed sample is somewhat brighter, so the additional magnification of the sheared systems may have figured more prominently, but not enough to account for an increase from 10\% shear prevalence to 62\%.}
	{The mock lens catalog has a magnitude limit on the third brightest lensed image and a minimum separation between the two most distant images. The observed sample is drawn from many different surveys, each with different magnitude and separation criteria, but the flux from the third brightest image and the largest separation (or perhaps the larger distance of the two brighter images from the third) are likely to have played a similarly dominant role in their selection. It is therefore not obvious how differences in selection effects would offset the mock and observed lenses from each other.}
	
	\section{Magnification bias}\label{sec:magbias}
	
	\added{While the magnification selection effects for our observed sample may not be different from those of the mock catalog, such effects \textit{do} affect the location of \textit{the boundary} between ellipticity and shear dominated systems.}
	
	\added{\citet{finch} show that the mean magnification for purely sheared systems is roughly a factor of two larger than for purely flattened systems. We suspect that this also applies to the individual images for a given hyperbola and ellipse. The unmagnified limiting magnitude for a pure shear system is therefore $0.75$ magnitudes fainter than for a pure ellipticity system.}
	
	\added{But the area of a pure shear caustic (and hence the cross section for quadruple lensing) is one quarter that of a pure ellipticity caustic. Therefore, the surface density of quasars must increase as $10^{0.8m}$ -- much more rapidly than observed at the relevant apparent magnitudes -- for the two effects to cancel. If they did exactly cancel, the opening angle between the dotted lines in Figures \ref{fig:gammavseta}a and \ref{fig:gammavseta}b would be $90^\circ$ rather than $45^\circ$. For a less steep rise in the number magnitude relation, the opening angle would be only somewhat larger than $45^\circ$. This would lead to a correspondingly stronger conclusion regarding shear dominance.}
	
	\section{Conclusion}
	
	Using the geometric properties of the \added{restricted} SIEP+XS model, we analyzed 39 observed quadruply lensed quasar systems with a single lensing galaxy. Comparing the observed galaxy center with the model, we estimated the shear and ellipticity components parallel to the effective quadrupole for each system. Using the deviation between the \replaced{true}{known input} and model galaxy position as well as the published uncertainty, we estimated the uncertainty of each decomposition and found that 15 systems out of 39 are reliably or provisionally shear-dominated while 11 systems are reliably or provisionally ellipticity-dominated. We also tested our method with the simulated mock \replaced{catalogue}{catalog} of \citet{oguri} and while the decompositions seem to be mostly consistent with the true shear and ellipticity values, the systematic difference between observed and simulated systems suggests that the effect of external shear has been underestimated in creating the mock \replaced{catalogue}{catalog}.
	
	\added{\citet{oguri} have made their code available and it would appear to be a straightfoward matter to adjust the parameters governing the mean ellipticity and mean shear. We would suggest that users of their code iterate on their input parameters so as to yield a selected median semi-ellipticity of $0.05$ and a selected median shear of $0.12$. Decreasing the input ellipticity and increasing the input shear will have opposite effects on the total number of selected systems.}
	
	\acknowledgements 
	
	We thank Drs. Philip Marshall and Masamune Oguri for comments on the
	manuscript. This research is based on observations made with the NASA/ESA Hubble Space Telescope obtained from the Space Telescope Science Institute, which is operated by the Association of Universities for Research in Astronomy, Inc., under NASA contract NAS5-26555, as part of programs HST-GO-15320 and HST-GO-15652.
	
	\clearpage
	\appendix
	\section{Geometric properties of the restricted SIEP+XS model}\label{app:geom}
	
	Using the lens equation (\ref{eq:lensequation}), we get the following equations for the coordinates of images $\mathbf{r} = (x,y)$:
	\begin{align}
		x-x_s &= -\gamma(x-x_s)+\frac{b(x-x_g)q_\text{pot}}{\sqrt{q_\text{pot}(x-x_g)^2+(y-y_g)^2/q_\text{pot}}},\\
		y-y_s &= \phantom{-}\gamma(y-y_s)+\frac{b(y-y_g)/q_\text{pot}}{\sqrt{q_\text{pot}(x-x_g)^2+(y-y_g)^2/q_\text{pot}}}.
	\end{align}
	
	Defining $t \equiv \sqrt{q_\text{pot}(x-x_g)^2+(y-y_g)^2/q_\text{pot}}$ gives
	\begin{align}
		x-x_s &= \frac{b(x-x_g)q_\text{pot}}{(1+\gamma)t},\\
		y-y_s &= \frac{b(y-y_g)/q_\text{pot}}{(1-\gamma)t}.   
	\end{align}
	
	We get the hyperbola equation (\ref{eq:hyperbola}) by dividing the two equations. To get the ellipse equation (\ref{eq:ellipse}), we notice that
	\begin{equation}
		\left(x-x_s\right)^2 + \left(\frac{1-\gamma}{1+\gamma}\right)^2 q_\text{pot}^2 (y-y_s)^2 = \frac{b^2q_\text{pot}^2 (x-x_g)^2}{(1+\gamma)^2 t^2} + \frac{b^2 (y-y_g)^2}{(1+\gamma)^2 t^2} = \frac{b^2q_\text{pot}}{(1+\gamma)^2}.
	\end{equation}
	
	
	\section{Equivalence of shear and ellipticity for image positions}\label{app:imageequivalence}
	
	To show that there are SIEP+XS lenses with different shear-ellipticity proportions that give the same image positions, it is sufficient to show that when we change shear and ellipticity of a given lens system, we can choose the other parameters ($b$, $x_g$, $y_g$) so that we get the same ellipse and hyperbola back again (because the four intersection points of the ellipse and hyperbola uniquely determine the image positions). For simplicity, let us fix the position of the source at $(x_s, y_s)$ and let us work in the aligned frame, with the origin at the center of the hyperbola (so the equation of the hyperbola is $xy = \text{const}$).
	
	Assuming we do not change the position of the source, it is necessary to have the same deflection ellipse to have the same image positions because an ellipse is generally uniquely defined by 4 points and its center. Therefore, when we are only changing $b$, $x_g$, and $y_g$, the axis ratio of the ellipse must stay the same:
	\begin{equation}
		q_\text{ell} = \left(\frac{1-\gamma}{1+\gamma}\right)q_\text{pot} = \text{const}.
	\end{equation}
	The right hand side of equation (\ref{eq:ellipse}) must also stay the same, which determines how $b$ should be changed:
	\begin{equation}
		\frac{b^2q_\text{pot}}{(1+\gamma)^2} = \text{const}.
	\end{equation}
	Conversely, it shows that when $q_\text{ell}$ is constrained to a fixed value while changing shear and ellipticity, we can always choose a new value for $b$ that keeps the deflection ellipse the same.
	
	To show that we can also choose parameters $x_g$ and $y_g$ which keep the hyperbola same, let us choose an arbitrary point $A=(x_A, y_A)$ on the same branch of the hyperbola as the source. Let us then choose the point $(x_g, y_g)$ on the hyperbola so that
	\begin{equation}
		\frac{(y_A-y_s)}{(x_A-x_s)}\frac{(x_A-x_g)}{(y_A-y_g)} = \left(\frac{1+\gamma}{1-\gamma}\right)\frac{1}{q_\text{pot}^2}.
	\end{equation}
	Note that we can always find such $(x_g,y_g)$ because the first factor is a constant and the limiting values of the second factor for a rectangular hyperbola are $0$ and $\infty$. Because $(x_A,y_A)$ satisfies equation (\ref{eq:hyperbola}) for the new system, points $(x_A, y_A)$, $(x_s, y_s)$, and $(x_g, y_g)$ are also on the new hyperbola. According to the inscribed angle theorem for hyperbolas, a rectangular hyperbola in the aligned frame is uniquely determined by three points:
	\begin{equation}
		\frac{(y-y_1)}{(x-x_1)}\frac{(x-x_2)}{(y-y_2)} = \frac{(y_3-y_1)}{(x_3-x_1)}\frac{(x_3-x_2)}{(y_3-y_2)},
	\end{equation}
	which shows that the new hyperbola coincides with the old one. Therefore, the image positions also stay the same.
	
	\added{To explicitly find $x_g$ and $y_g$, we rewrite equation (\ref{eq:hyperbola}) as}
	\begin{equation}
		\frac{y-y_s}{y-y_g} = \left(\frac{1+\gamma}{1-\gamma}\right)\frac{1}{q_\text{pot}^2}\frac{x-x_s}{x-x_g}.
	\end{equation}
	
	\added{Paralleling the argument of \citet{wynne}, we consider the limit $x \to 0$ and $y \to \infty$ (note that $xy=\text{const}$). In this case, the left hand side approaches $1$ and $(x-x_s)/(x-x_g) \to x_s/x_g$, so we obtain}
	\begin{equation}
		x_g = \left(\frac{1+\gamma}{1-\gamma}\right)\frac{1}{q_\text{pot}^2}x_s.
	\end{equation}
	
	\added{Similarly, by considering the limit $y \to 0$ and $x \to \infty$, we obtain}
	\begin{equation}
		y_g = \left(\frac{1-\gamma}{1+\gamma}\right) q_\text{pot}^2 y_s.
	\end{equation}
	
	
	\section{Formula for the area of caustic}\label{app:caustic}
	
	We find the area of the \replaced{astroid}{astroidal} caustic following the argument of \citet{finch} for the SIS+XS model.
	The inverse magnification of a lensed image is given by
	\begin{equation}
		\mu^{-1} = \left(1-\frac{\partial^2\psi}{\partial x^2}\right)\left(1-\frac{\partial^2\psi}{\partial y^2}\right) - \left(\frac{\partial^2 \psi}{\partial x \partial y}\right)^2.
	\end{equation}
	Taking $(x_g, y_g) = (0,0)$ for simplicity and defining $t \equiv \sqrt{q_\text{pot}x^2+y^2/q_\text{pot}}$, we use the lensing potential given by equation (\ref{eq:psi}) to obtain
	\begin{align}
		\mu^{-1} &= \left(1+\gamma - \frac{by^2}{t^3}\right)\left(1-\gamma - \frac{bx^2}{t^3}\right) - \frac{b^2x^2y^2}{t^6}, \\
		\mu^{-1} &= (1-\gamma^2) -(1+\gamma)\frac{bx^2}{t^3} - (1-\gamma)\frac{by^2}{t^3}.
	\end{align}
	
	The critical line is the locus of points where the inverse magnification $\mu^{-1}$ is 0 \citep{finch}, therefore the equation for the critical line $(x_c, y_c)$ is
	\begin{align}
		(1-\gamma^2)t^3 &= (1+\gamma)bx_c^2 + (1-\gamma)by_c^2, \\
		\left(q_\text{pot}x_c^2 + \frac{y_c^2}{q_\text{pot}}\right)^{3/2} &= \frac{bx_c^2}{1-\gamma} + \frac{by_c^2}{1+\gamma}.
	\end{align}
	We can get parametric equations for $x_c$ and $y_c$ in terms of $\theta_c$ by substituting $x_c = r_c\cos\theta_c$ and $y_c = r_c\sin\theta_c$:
	\begin{equation}
		r_c\left(q_\text{pot}\cos^2\theta_c +
		\frac{\sin^2\theta_c}{q_\text{pot}}\right)^{3/2} = \frac{b\cos^2\theta_c}{1-\gamma} + \frac{b\sin^2\theta_c}{1+\gamma},
	\end{equation}
	\begin{equation}
		r_c = b\left(\frac{\cos^2\theta_c}{1-\gamma} + \frac{\sin^2\theta_c}{1+\gamma}\right)\left(q_\text{pot}\cos^2\theta_c +
		\frac{\sin^2\theta_c}{q_\text{pot}}\right)^{-3/2},
	\end{equation}
	hence
	\begin{align}
		x_c &= b\cos\theta_c\left(\frac{\cos^2\theta_c}{1-\gamma} + \frac{\sin^2\theta_c}{1+\gamma}\right)\left(q_\text{pot}\cos^2\theta_c +
		\frac{\sin^2\theta_c}{q_\text{pot}}\right)^{-3/2},\\
		y_c &= b\sin\theta_c\left(\frac{\cos^2\theta_c}{1-\gamma} + \frac{\sin^2\theta_c}{1+\gamma}\right)\left(q_\text{pot}\cos^2\theta_c +
		\frac{\sin^2\theta_c}{q_\text{pot}}\right)^{-3/2}.
	\end{align}
	We obtain the equation for the \replaced{astroid}{astroidal} caustic $(x_a, y_a)$ by reverse mapping with equation (\ref{eq:lensequation}), that is
	\begin{equation}
		\mathbf{r}_a = \mathbf{r}_c - \boldsymbol{\nabla}\psi(\mathbf{r}_c).
	\end{equation}
	Therefore
	\begin{align}
		x_a &= x_c + \gamma(x_c-x_s) - \frac{bx_cq_\text{pot}}{t},\\
		y_a &= y_c - \gamma(y_c-y_s) - \frac{by_c/q_\text{pot}}{t}.
	\end{align}
	We notice that $x_s$ and $y_s$ only shift the caustic and do not change its shape, so we can take them to be 0 without loss of generality. After simplifying we get
	\begin{align}
		x_a &= \phantom{-}\frac{b\left(1+\gamma - (1-\gamma)q_\text{pot}^2\right)}{\left(q_\text{pot}\cos^2\theta_c + \left(1/q_\text{pot}\right)\sin^2\theta_c\right)^{3/2}}\frac{\cos^3 \theta_c}{(1-\gamma)},\\
		y_a &= -\frac{b\left(1+\gamma - (1-\gamma)q_\text{pot}^2\right)}{\left(q_\text{pot}\cos^2\theta_c + \left(1/q_\text{pot}\right)\sin^2\theta_c\right)^{3/2}}\frac{\sin^3 \theta_c}{(1+\gamma)q_\text{pot}^2}.
	\end{align}
	\added{These are similar, but not identical to the parametric equations for an astroid, for which the coefficients of the $\sin^3$ and $\cos^3$ terms would be identical. For the case of pure shear, they give a stretched astroid.} The area of the \replaced{astroid}{astroidal} caustic is
	\begin{align}
		A_a = 4\left|\int_0^{x_\text{max}}y_a\,\mathrm{d}x_a\right| &= \frac{12 b^2 \left(\left(1-q_\text{pot}^2\right)+\left(1+q_\text{pot}^2\right)\gamma\right)^2}{(1-\gamma^2)q_\text{pot}^3}\int_0^{\pi/2}\frac{\cos^2\theta_c \sin^4 \theta_c}{\left(q_\text{pot}\cos^2\theta_c + \left(1/q_\text{pot}\right)\sin^2\theta_c\right)^4}\,\mathrm{d}\theta_c \\\nonumber
		&= \frac{3\pi b^2 \left(\left(1-q_\text{pot}^2\right)+\left(1+q_\text{pot}^2\right)\gamma\right)^2}{8(1-\gamma^2)q_\text{pot}^2}.
	\end{align}
	
	
	\section{Fitting of hyperbola and ellipse}\label{app:fitting}
	
	The first step in our determination of the relative contributions of flattening and shear to a given quadruply lensed quasar is to find the hyperbola and ellipse that give a ``best fit'' to the four image positions, subject to the constraint that the center of the ellipse
	must lie on the hyperbola and that the major and minor axes of the ellipse are aligned with the asymptotes of the hyperbola.
	
	\citet{wynne} and \citet{schechterwynne} describe two alternative schemes for obtaining such a best fit. The first of these has the shortcoming that it produces models in which the ellipse intersects only the primary branch of the hyperbola (a two image lens model). The latter has the shortcoming that while two of the observed images lie precisely where the ellipse intersects the secondary branch of the hyperbola, the other two may lie far from the primary branch. 
	
	We have adopted an intermediate scheme. In our first implementation we minimized the sum of the squared distances of the observed images from the hyperbola and from the ellipse (rather than from their points of intersection). This worked poorly in a few cases where the ellipse barely intersects the secondary branch of the hyperbola and is almost tangent to it. The results were more satisfactory when we weighted images close to each other more heavily. Specifically, we multiplied the sum of squared distances of each image by
	\begin{equation}
		w_i = \min\left(10^{0.05\left(d_\text{max}/d_\text{min}\right)^2}, 1000\right),
	\end{equation}
	where $d_\text{max}$ is the distance of the farthest image and $d_\text{min}$ is the distance of the closest image among the other three. 
	
	Had we sought perfection, we might have, at the cost of additional coding and CPU time, minimized the distances of the images from the points of intersection, but our scheme suffices for the present task.
	
	\clearpage
	\section{Relation between ellipticity of mass distribution and ellipticity of lensing potential}\label{app:potmassrelation}
	
	\citet{keeton1998} have shown that for a singular isothermal ellipsoid with the surface mass density given by
	\begin{equation}
		\Sigma/\Sigma_\text{cr} = \frac{b_I}{2\sqrt{q_\text{mass}^2 x^2 + y^2}},
	\end{equation}
	the deflection angle of each coordinate is
	\begin{align}
		\alpha_{x,\text{mass}} &= \frac{b_I}{\sqrt{1-q_\text{mass}^2}}\tan^{-1}\left(\frac{x\sqrt{1-q_\text{mass}^2 }}{\sqrt{q_\text{mass}^2 x^2+y^2}}\right),\\
		\alpha_{y,\text{mass}} &= \frac{b_I}{\sqrt{1-q_\text{mass}^2}}\tanh^{-1}\left(\frac{y\sqrt{1-q_\text{mass}^2 }}{\sqrt{q_\text{mass}^2 x^2+y^2}}\right),
	\end{align}
	where $b_I$ is a scale factor (not necessarily equal to $b$). For our model, using equation (\ref{eq:lensequation}) and taking $\gamma = 0$, $(x_g,y_g)=(0,0)$ gives
	\begin{align}
		\alpha_{x,\text{pot}} &= \frac{bxq_\text{pot}}{\sqrt{q_\text{pot}x^2+y^2/q_\text{pot}}},\\
		\alpha_{y,\text{pot}} &= \frac{by/q_\text{pot}}{\sqrt{q_\text{pot}x^2+y^2/q_\text{pot}}}.
	\end{align}
	
	Although the models give different results for large ellipticities, we can find an approximate mapping between $q_\text{mass}$ and $q_\text{pot}$ by setting the deflection angles equal in special cases where one of the coordinates is 0. Setting $y=0$ and $\alpha_{x,\text{pot}} = \alpha_{x,\text{mass}}$ gives
	\begin{align}
		bq_\text{pot}^{1/2} = \frac{b_I}{\sqrt{1-q_\text{mass}^2}}\tan^{-1}\left(\frac{\sqrt{1-q_\text{mass}^2 }}{q_\text{mass}}\right).
	\end{align}
	Setting $x=0$ and $\alpha_{y,\text{pot}} = \alpha_{y,\text{mass}}$ gives
	\begin{align}
		bq_\text{pot}^{-1/2} = \frac{b_I}{\sqrt{1-q_\text{mass}^2}}\tanh^{-1}\left(\sqrt{1-q_\text{mass}^2 }\right).
	\end{align}
	We obtain the relation by dividing the two equations:
	\begin{equation}\label{eq:qpotqmassappendix}
		q_\text{pot} = \frac{\tan^{-1}\left(\sqrt{1-q_\text{mass}^2}/q_\text{mass}\right)}{\tanh^{-1}\left(\sqrt{1-q_\text{mass}^2}\right)}.
	\end{equation}
	
	By plotting the function, we can note that $q_\text{pot} = q_\text{mass}^{1/3}$ is a good approximation of this relation. Nevertheless, we used equation (\ref{eq:qpotqmassappendix}) to convert the axis ratio.
	
	\clearpage
	\section{Table of image and galaxy positions}\label{app:positions}
	
	\startlongtable
	\begin{deluxetable}{lDDDDDDDDr@{$\,\pm\,$}lr@{$\,\pm\,$}ll}
		\centerwidetable
		\tabletypesize{\scriptsize}
		\tablecaption{Known quadruply lensed quasar systems and the relative observed positions of the images and the lensing galaxy \label{tab:systems}}
		\tablehead{
			\colhead{} &
			\colhead{} &  \colhead{} &  \colhead{} &  \colhead{} &
			\colhead{} &  \colhead{} &  \colhead{} &  \colhead{} &
			\colhead{} &  \colhead{} &  \colhead{} &  \colhead{} &
			\colhead{} &  \colhead{} &  \colhead{} &  \colhead{} &
			\colhead{}  \\
			\colhead{} & \multicolumn{4}{c}{Image A} & \multicolumn{4}{c}{Image B} & \multicolumn{4}{c}{Image C} & \multicolumn{4}{c}{Image D} & \multicolumn{4}{c}{Galaxy} & \colhead{} \\  
			\colhead{System name} & \twocolhead{$\Delta \alpha \cos \delta$} & \twocolhead{$\Delta \delta$} & \twocolhead{$\Delta \alpha \cos \delta$} & \twocolhead{$\Delta \delta$} & \twocolhead{$\Delta \alpha \cos \delta$} & \twocolhead{$\Delta \delta$} & \twocolhead{$\Delta \alpha \cos \delta$} & \twocolhead{$\Delta \delta$} & \multicolumn{2}{c}{$\Delta \alpha \cos \delta$} & \multicolumn{2}{c}{$\Delta \delta$} & \colhead{Ref.} \\
			\colhead{} & \twocolhead{(arcsec)} & \twocolhead{(arcsec)} & \twocolhead{(arcsec)} & \twocolhead{(arcsec)} & \twocolhead{(arcsec)} & \twocolhead{(arcsec)} & \twocolhead{(arcsec)} & \twocolhead{(arcsec)} & \multicolumn{2}{c}{(arcsec)} & \multicolumn{2}{c}{(arcsec)} & \colhead{}
		}
		\decimals
		\startdata
		J0029-3814 & $0.000$ & $0.000$ & $2.132$ & $-0.585$ & $0.656$ & $0.313$ & $0.384$ & $-0.558$ & $0.850$ & $0.003$ & $-0.223$ & $0.003$ & 1 \\
		J0030-1525 & $0.000$ & $0.000$ & $-0.021$ & $-1.773$ & $-1.641$ & $-0.861$ & $0.204$ & $-0.270$ & $-0.872$ & $0.002$ & $-0.832$ & $0.002$ & 1 \\
		PS J0147+4630 & $0.155$ & $2.051$ & $1.327$ & $1.642$ & $-1.084$ & $1.958$ & $-0.186$ & $-1.170$ & $0.000$ & $0.0002$ & $0.000$ & $0.0001$ & 2 \\
		SDSS J0248+1913 & $-0.647$ & $-0.204$ & $-0.505$ & $0.629$ & $0.351$ & $-0.821$ & $0.401$ & $0.590$ & $0.000$ & $0.001$ & $0.000$ & $0.001$ & 2 \\
		ATLAS J0259-1635 & $0.683$ & $-0.303$ & $0.357$ & $0.571$ & $-0.801$ & $0.253$ & $-0.043$ & $-0.700$ & $0.000$ & $0.003$ & $0.000$ & $0.001$ & 2 \\
		DES J0405-3308 & $0.694$ & $-0.238$ & $-0.375$ & $-0.561$ & $0.344$ & $0.603$ & $-0.525$ & $0.454$ & $0.000$ & $0.001$ & $0.000$ & $0.001$ & 2 \\
		DES J0420-4037 & $-0.697$ & $-0.350$ & $-0.457$ & $0.683$ & $0.711$ & $-0.568$ & $0.172$ & $0.788$ & $0.000$ & $0.001$ & $0.000$ & $0.001$ & 2 \\
		HE0435-1223 & $0.000$ & $0.000$ & $-1.476$ & $0.553$ & $-2.467$ & $-0.603$ & $-0.939$ & $-1.614$ & $-1.165$ & $0.003$ & $-0.573$ & $0.003$ & 3 \\
		J0530-3730 & $0.000$ & $0.000$ & $-0.367$ & $-0.562$ & $-0.127$ & $-0.100$ & $0.615$ & $-0.841$ & $0.336$ & $0.015$ & $-0.638$ & $0.015$ & 1 \\
		J0659+16 & $0.000$ & $0.000$ & $-4.766$ & $-2.210$ & $-1.056$ & $1.005$ & $-0.088$ & $-1.877$ & $-1.884$ & $0.02$ & $-1.041$ & $0.02$ & 1 \\
		B0712+472 & $0.000$ & $0.000$ & $0.056$ & $-0.156$ & $0.812$ & $-0.663$ & $1.174$ & $0.459$ & $0.785$ & $0.003$ & $0.142$ & $0.003$ & 4 \\
		HS0810+2554 & $0.000$ & $0.000$ & $0.087$ & $-0.163$ & $0.774$ & $-0.257$ & $0.610$ & $0.579$ & $0.451$ & $0.012$ & $0.143$ & $0.005$ & 3 \\
		RXJ0911+0551 & $0.000$ & $0.000$ & $0.260$ & $0.406$ & $-0.018$ & $0.960$ & $-2.972$ & $0.792$ & $-0.698$ & $0.004$ & $0.512$ & $0.005$ & 3 \\
		SDSS0924+0219 & $0.000$ & $0.000$ & $0.061$ & $-1.805$ & $-0.968$ & $-0.676$ & $0.536$ & $-0.427$ & $-0.182$ & $0.003$ & $-0.859$ & $0.003$ & 3 \\
		J1042+1641 & $0.000$ & $0.000$ & $-0.152$ & $-0.566$ & $-0.813$ & $-0.909$ & $-1.592$ & $0.541$ & $-0.792$ & $0.013$ & $-0.076$ & $0.003$ & 5 \\
		PG1115+080 & $1.328$ & $-2.034$ & $1.477$ & $-1.576$ & $-0.341$ & $-1.961$ & $0.000$ & $0.000$ & $0.381$ & $0.003$ & $-1.344$ & $0.003$ & 3 \\
		RXJ1131-1231 & $0.588$ & $1.120$ & $0.618$ & $2.307$ & $0.000$ & $0.000$ & $-2.517$ & $1.998$ & $-1.444$ & $0.008$ & $1.706$ & $0.006$ & 3 \\
		J1131-4419 & $0.000$ & $0.000$ & $-0.334$ & $0.342$ & $-1.628$ & $-0.098$ & $-0.689$ & $-1.182$ & $-0.763$ & $0.002$ & $-0.403$ & $0.002$ & 1 \\
		J1134-2103 & $0.000$ & $0.000$ & $-2.676$ & $-2.528$ & $-0.729$ & $-1.757$ & $-1.976$ & $-0.390$ & $-1.471$ & $0.003$ & $-0.970$ & $0.003$ & 1 \\
		SDSS1138+0314 & $0.000$ & $0.000$ & $-0.103$ & $0.979$ & $-1.184$ & $0.812$ & $-0.698$ & $-0.056$ & $-0.474$ & $0.003$ & $0.534$ & $0.003$ & 3 \\
		SDSS J1251+2935 & $0.346$ & $-0.616$ & $0.707$ & $-0.257$ & $0.637$ & $0.335$ & $-1.080$ & $0.319$ & $0.000$ & $0.0005$ & $0.000$ & $0.0005$ & 2 \\
		HST12531-2914 & $-0.737$ & $-0.011$ & $0.605$ & $-0.339$ & $-0.140$ & $-0.519$ & $0.166$ & $0.454$ & $0.000$ & $0.023$ & $0.000$ & $0.016$ & 3 \\
		SDSS J1330+1810 & $0.000$ & $0.000$ & $-0.414$ & $-0.011$ & $-1.248$ & $1.168$ & $0.244$ & $1.579$ & $-0.221$ & $0.008$ & $0.966$ & $0.0015$ & 6 \\
		HST14113+5211 & $-1.095$ & $0.368$ & $-0.118$ & $-0.551$ & $1.132$ & $-0.036$ & $0.268$ & $0.695$ & $0.000$ & $0.003$ & $0.000$ & $0.007$ & 3 \\
		H1413+117 & $0.000$ & $0.000$ & $0.744$ & $0.168$ & $-0.492$ & $0.713$ & $0.354$ & $1.040$ & $0.142$ & $0.003$ & $0.561$ & $0.003$ & 3 \\
		HST14176+5226 & $1.288$ & $-1.175$ & $0.880$ & $0.879$ & $-0.792$ & $1.332$ & $-0.808$ & $-0.794$ & $0.000$ & $0.008$ & $0.000$ & $0.01$ & 3 \\
		B1422+231 & $0.385$ & $0.317$ & $0.000$ & $0.000$ & $-0.336$ & $-0.750$ & $0.948$ & $-0.802$ & $0.742$ & $0.003$ & $-0.656$ & $0.004$ & 3 \\
		SDSS J1433+6007 & $-0.941$ & $2.058$ & $-0.943$ & $-1.691$ & $-1.721$ & $-0.083$ & $1.075$ & $-0.138$ & $0.000$ & $0.002$ & $0.000$ & $0.003$ & 2 \\
		J1537-30 & $0.000$ & $0.000$ & $2.842$ & $-1.648$ & $0.847$ & $-1.964$ & $2.100$ & $0.123$ & $1.404$ & $0.006$ & $-0.889$ & $0.006$ & 1 \\
		PS J1606-2333 & $0.833$ & $0.373$ & $-0.793$ & $-0.223$ & $0.040$ & $-0.541$ & $-0.296$ & $0.524$ & $0.000$ & $0.001$ & $0.000$ & $0.001$ & 2 \\
		J1721+8842 & $0.000$ & $0.000$ & $-1.331$ & $1.401$ & $-0.236$ & $-2.348$ & $-3.319$ & $-1.144$ & $-1.881$ & $0.004$ & $-0.740$ & $0.004$ & 1 \\
		J1817+27 & $0.000$ & $0.000$ & $1.257$ & $-1.283$ & $1.263$ & $0.159$ & $-0.095$ & $-0.803$ & $0.694$ & $0.003$ & $-0.440$ & $0.003$ & 1 \\
		WFI2026-4536 & $0.163$ & $-1.428$ & $0.416$ & $-1.214$ & $0.000$ & $0.000$ & $-0.572$ & $-1.042$ & $-0.074$ & $0.012$ & $-0.798$ & $0.008$ & 3 \\
		DES J2038-4008 & $-1.482$ & $0.499$ & $0.834$ & $-1.212$ & $-0.688$ & $-1.182$ & $0.704$ & $0.864$ & $0.000$ & $0.001$ & $0.000$ & $0.001$ & 2 \\
		B2045+265 & $0.000$ & $0.000$ & $-0.134$ & $-0.248$ & $-0.288$ & $-0.790$ & $1.629$ & $-1.006$ & $1.108$ & $0.001$ & $-0.807$ & $0.001$ & 7 \\
		J2100-4452 & $0.000$ & $0.000$ & $-0.422$ & $0.332$ & $2.020$ & $0.680$ & $0.421$ & $2.211$ & $0.611$ & $0.003$ & $1.141$ & $0.003$ & 1 \\
		J2145+6345 & $0.000$ & $0.000$ & $0.316$ & $0.579$ & $-1.528$ & $-0.354$ & $-1.061$ & $1.318$ & $-0.733$ & $0.008$ & $0.693$ & $0.008$ & 1 \\
		J2205-3727 & $0.000$ & $0.000$ & $0.443$ & $-0.603$ & $1.642$ & $0.122$ & $0.295$ & $0.557$ & $0.728$ & $0.003$ & $0.016$ & $0.003$ & 1 \\
		WISE J2344-3056 & $-0.452$ & $0.179$ & $0.133$ & $0.530$ & $-0.212$ & $-0.478$ & $0.421$ & $-0.140$ & $0.000$ & $0.001$ & $0.000$ & $0.001$ & 2 \\
		\enddata
		\tablerefs{(1) Authors' preliminary measurements; authoritative measurements in Schmidt et al. (to be submitted); (2) \citet{shajib}; (3) \citet{castles}; (4) \citet{hsueh}; (5) \citet{glikman}; (6) \citet{rusu}; (7) \citet{mckean}}
	\end{deluxetable}
	
	
	\bibliography{WMQLQQ}{}

\begin{thebibliography}{}
\expandafter\ifx\csname natexlab\endcsname\relax\def\natexlab#1{#1}\fi
\providecommand{\url}[1]{\href{#1}{#1}}
\providecommand{\dodoi}[1]{doi:~\href{http://doi.org/#1}{\nolinkurl{#1}}}
\providecommand{\doeprint}[1]{\href{http://ascl.net/#1}{\nolinkurl{http://ascl.net/#1}}}
\providecommand{\doarXiv}[1]{\href{https://arxiv.org/abs/#1}{\nolinkurl{https://arxiv.org/abs/#1}}}

\bibitem[{{An}(2005)}]{jinan}
{An}, J.~H. 2005, \mnras, 356, 1409, \dodoi{10.1111/j.1365-2966.2004.08581.x}

\bibitem[{{Bourassa} \& {Kantowski}(1975)}]{bourassa1975}
{Bourassa}, R.~R., \& {Kantowski}, R. 1975, \apj, 195, 13,
  \dodoi{10.1086/153300}

\bibitem[{{Collett}(2015)}]{collett}
{Collett}, T.~E. 2015, \apj, 811, 20, \dodoi{10.1088/0004-637X/811/1/20}

\bibitem[{{Delchambre} {et~al.}(2019){Delchambre}, {Krone-Martins}, {Wertz},
  {Ducourant}, {Galluccio}, {Kl{\"u}ter}, {Mignard}, {Teixeira}, {Djorgovski},
  {Stern}, {Graham}, {Surdej}, {Bastian}, {Wambsganss}, {Le Campion}, \&
  {Slezak}}]{delchambre}
{Delchambre}, L., {Krone-Martins}, A., {Wertz}, O., {et~al.} 2019, \aap, 622,
  A165, \dodoi{10.1051/0004-6361/201833802}

\bibitem[{{Finch} {et~al.}(2002){Finch}, {Carlivati}, {Winn}, \&
  {Schechter}}]{finch}
{Finch}, T.~K., {Carlivati}, L.~P., {Winn}, J.~N., \& {Schechter}, P.~L. 2002,
  \apj, 577, 51, \dodoi{10.1086/342163}

\bibitem[{{Glikman} {et~al.}(2018){Glikman}, {Rusu}, {Djorgovski}, {Graham},
  {Stern}, {Urrutia}, {Lacy}, \& {O'Meara}}]{glikman}
{Glikman}, E., {Rusu}, C.~E., {Djorgovski}, S.~G., {et~al.} 2018, arXiv
  e-prints, arXiv:1807.05434.
\newblock \doarXiv{1807.05434}

\bibitem[{{Goldstein} {et~al.}(2019){Goldstein}, {Nugent}, \&
  {Goobar}}]{goldstein}
{Goldstein}, D.~A., {Nugent}, P.~E., \& {Goobar}, A. 2019, \apjs, 243, 6,
  \dodoi{10.3847/1538-4365/ab1fe0}

\bibitem[{{Gorenstein} {et~al.}(1988){Gorenstein}, {Falco}, \&
  {Shapiro}}]{gorenstein}
{Gorenstein}, M.~V., {Falco}, E.~E., \& {Shapiro}, I.~I. 1988, \apj, 327, 693,
  \dodoi{10.1086/166226}

\bibitem[{{Holder} \& {Schechter}(2003)}]{holderschechter}
{Holder}, G.~P., \& {Schechter}, P.~L. 2003, \apj, 589, 688,
  \dodoi{10.1086/374688}

\bibitem[{{Hsueh} {et~al.}(2017){Hsueh}, {Oldham}, {Spingola}, {Vegetti},
  {Fassnacht}, {Auger}, {Koopmans}, {McKean}, \& {Lagattuta}}]{hsueh}
{Hsueh}, J.~W., {Oldham}, L., {Spingola}, C., {et~al.} 2017, \mnras, 469, 3713,
  \dodoi{10.1093/mnras/stx1082}

\bibitem[{{Huterer} {et~al.}(2005){Huterer}, {Keeton}, \& {Ma}}]{huterer}
{Huterer}, D., {Keeton}, C.~R., \& {Ma}, C.-P. 2005, \apj, 624, 34,
  \dodoi{10.1086/429153}

\bibitem[{{Kassiola} \& {Kovner}(1993)}]{kassiolakovner}
{Kassiola}, A., \& {Kovner}, I. 1993, \apj, 417, 450, \dodoi{10.1086/173325}

\bibitem[{{Keeton} \& {Kochanek}(1998)}]{keeton1998}
{Keeton}, C.~R., \& {Kochanek}, C.~S. 1998, \apj, 495, 157,
  \dodoi{10.1086/305272}

\bibitem[{{Keeton} {et~al.}(1997){Keeton}, {Kochanek}, \& {Seljak}}]{kks}
{Keeton}, C.~R., {Kochanek}, C.~S., \& {Seljak}, U. 1997, \apj, 482, 604,
  \dodoi{10.1086/304172}

\bibitem[{Kochanek {et~al.}(1999)Kochanek, Falco, Impey, Lehar, McLeod, \&
  Rix}]{castles}
Kochanek, C.~S., Falco, E.~E., Impey, C., {et~al.} 1999, {{CASTLES
  Gravitational Lens Data Base}}.
\newblock \url{https://www.cfa.harvard.edu/castles/}

\bibitem[{{McKean} {et~al.}(2007){McKean}, {Koopmans}, {Flack}, {Fassnacht},
  {Thompson}, {Matthews}, {Bland ford}, {Readhead}, \& {Soifer}}]{mckean}
{McKean}, J.~P., {Koopmans}, L.~V.~E., {Flack}, C.~E., {et~al.} 2007, \mnras,
  378, 109, \dodoi{10.1111/j.1365-2966.2007.11744.x}

\bibitem[{{Oguri} \& {Marshall}(2010)}]{oguri}
{Oguri}, M., \& {Marshall}, P.~J. 2010, \mnras, 405, 2579,
  \dodoi{10.1111/j.1365-2966.2010.16639.x}

\bibitem[{{Refsdal}(1964)}]{refsdal}
{Refsdal}, S. 1964, \mnras, 128, 307, \dodoi{10.1093/mnras/128.4.307}

\bibitem[{{Rusu} {et~al.}(2016){Rusu}, {Oguri}, {Minowa}, {Iye}, {Inada},
  {Oya}, {Kayo}, {Hayano}, {Hattori}, {Saito}, {Ito}, {Pyo}, {Terada},
  {Takami}, \& {Watanabe}}]{rusu}
{Rusu}, C.~E., {Oguri}, M., {Minowa}, Y., {et~al.} 2016, \mnras, 458, 2,
  \dodoi{10.1093/mnras/stw092}

\bibitem[{{Schechter} \& {Wynne}(2019)}]{schechterwynne}
{Schechter}, P.~L., \& {Wynne}, R.~A. 2019, \apj, 876, 9,
  \dodoi{10.3847/1538-4357/ab1258}

\bibitem[{{Schneider} \& {Sluse}(2014)}]{schneidersluse}
{Schneider}, P., \& {Sluse}, D. 2014, \aap, 564, A103,
  \dodoi{10.1051/0004-6361/201322106}

\bibitem[{{Shajib} {et~al.}(2019){Shajib}, {Birrer}, {Treu}, {Auger},
  {Agnello}, {Anguita}, {Buckley-Geer}, {Chan}, {Collett}, {Courbin},
  {Fassnacht}, {Frieman}, {Kayo}, {Lemon}, {Lin}, {Marshall}, {McMahon},
  {More}, {Morgan}, {Motta}, {Oguri}, {Ostrovski}, {Rusu}, {Schechter},
  {Shanks}, {Suyu}, {Meylan}, {Abbott}, {Allam}, {Annis}, {Avila}, {Bertin},
  {Brooks}, {Carnero Rosell}, {Carrasco Kind}, {Carretero}, {Cunha}, {da
  Costa}, {De Vicente}, {Desai}, {Doel}, {Flaugher}, {Fosalba},
  {Garc{\'\i}a-Bellido}, {Gerdes}, {Gruen}, {Gruendl}, {Gutierrez}, {Hartley},
  {Hollowood}, {Hoyle}, {James}, {Kuehn}, {Kuropatkin}, {Lahav}, {Lima},
  {Maia}, {March}, {Marshall}, {Melchior}, {Menanteau}, {Miquel}, {Plazas},
  {Sanchez}, {Scarpine}, {Sevilla-Noarbe}, {Smith}, {Soares-Santos},
  {Sobreira}, {Suchyta}, {Swanson}, {Tarle}, \& {Walker}}]{shajib}
{Shajib}, A.~J., {Birrer}, S., {Treu}, T., {et~al.} 2019, \mnras, 483, 5649,
  \dodoi{10.1093/mnras/sty3397}

\bibitem[{{Treu} \& {Marshall}(2016)}]{treumarshall}
{Treu}, T., \& {Marshall}, P.~J. 2016, \aapr, 24, 11,
  \dodoi{10.1007/s00159-016-0096-8}

\bibitem[{{Witt}(1996)}]{witt1996}
{Witt}, H.~J. 1996, \apjl, 472, L1, \dodoi{10.1086/310358}

\bibitem[{{Witt} \& {Mao}(2000)}]{wittmao}
{Witt}, H.~J., \& {Mao}, S. 2000, \mnras, 311, 689,
  \dodoi{10.1046/j.1365-8711.2000.03122.x}

\bibitem[{{Wong} {et~al.}(2011){Wong}, {Keeton}, {Williams}, {Momcheva}, \&
  {Zabludoff}}]{wong}
{Wong}, K.~C., {Keeton}, C.~R., {Williams}, K.~A., {Momcheva}, I.~G., \&
  {Zabludoff}, A.~I. 2011, \apj, 726, 84, \dodoi{10.1088/0004-637X/726/2/84}

\bibitem[{{Wynne} \& {Schechter}(2018)}]{wynne}
{Wynne}, R.~A., \& {Schechter}, P.~L. 2018, arXiv e-prints, arXiv:1808.06151.
\newblock \doarXiv{1808.06151}

\bibitem[{{Zahedy} {et~al.}(2016){Zahedy}, {Chen}, {Rauch}, {Wilson}, \&
  {Zabludoff}}]{zahedy}
{Zahedy}, F.~S., {Chen}, H.-W., {Rauch}, M., {Wilson}, M.~L., \& {Zabludoff},
  A. 2016, \mnras, 458, 2423, \dodoi{10.1093/mnras/stw484}

\bibitem[{{Zhao} \& {Pronk}(2001)}]{zhaopronk}
{Zhao}, H., \& {Pronk}, D. 2001, \mnras, 320, 401,
  \dodoi{10.1046/j.1365-8711.2001.03844.x}

\end{thebibliography}
	\bibliographystyle{aasjournal}
	
	\listofchanges
\end{document}